# A Hierarchical conv-LSTM and LLM Integrated Model for Holistic Stock Forecasting

Arya Chakraborty

*Dept. of Computer Science and Engineering*

*Birla Institute of Technology, Mesra*

Ranchi, India

btech10196.23@bitmesra.ac.in (ORCID: 0009-0002-3590-2891)

Auhona Basu

*Dept. of Computer Engineering*

*York University*

Toronto, Canada

auhona03@my.yorku.ca (ORCID: 0009-0002-4131-820X)

**Abstract:** The financial domain presents a complex environment for stock market prediction, characterized by volatile patterns and the influence of multifaceted data sources. Traditional models have leveraged either Convolutional Neural Networks (CNN) for spatial feature extraction or Long Short-Term Memory (LSTM) networks for capturing temporal dependencies, with limited integration of external textual data. This paper proposes a novel Two-Level Conv-LSTM Neural Network integrated with a Large Language Model (LLM) for comprehensive stock advising. The model harnesses the strengths of Conv-LSTM for analyzing time-series data and LLM for processing and understanding textual information from financial news, social media, and reports. In the first level, convolutional layers are employed to identify local patterns in historical stock prices and technical indicators, followed by LSTM layers to capture the temporal dynamics. The second level integrates the output with an LLM that analyzes sentiment and contextual information from textual data, providing a holistic view of market conditions. The combined approach aims to improve prediction accuracy and provide contextually rich stock advising.

## I. INTRODUCTION

Spatial data in stock markets refers to geographical factors and their relationship to financial activities and participants across the globe. By analysing trading volumes and the geographical distribution of market players, investors can gain insights into regional economic conditions and their impact on local share prices. Spatial analysis helps identify geographical risks, such as political instability or natural disasters, and provides an understanding of how regional regulations affect stock performance. This understanding enables investors to develop strategies tailored to specific regions and make more informed decisions that enhance their investment portfolios.

Temporal data in stock markets, on the other hand, involves monitoring changes in stock prices, trading volumes, and other market indicators over time, a process known as time-series analysis. Temporal data allows investors to observe trends, fluctuations, and patterns in stock performance, which are critical for making decisions and formulating strategies. For instance, by analysing historical prices, investors can predict future price movements based on long-term trends and short-term fluctuations. Temporal data is also essential for volatility analysis, where changes in market volatility are tracked over specific periods to gauge market stability and investor sentiment. Additionally, temporal data can be used for event impact analysis, helping investors understand how events like earnings announcements, economic reports, or geopolitical events have influenced stock prices over time. The continuous monitoring and analysis of temporal data can alert investors to potential market shifts, allowing them to adjust their strategies accordingly and optimize their investment outcomes.

The integration of spatial and temporal data in stock markets provides a comprehensive understanding of how geographical and time-related factors interact to influence market dynamics. Spatiotemporal data combines spatial information, such as the geographical locations of companies and economic activities, with temporal data on historical price trends and trading patterns. This integrated analysis enables investors to build a fuller understanding of market behaviour by examining how regional markets react to global events, political changes, natural phenomena, or economic policies, and how these reactions vary over time. For instance, spatiotemporal analysis can reveal variations in investor behaviour between different regions or significant periods, enabling more accurate predictive models and more effective investment strategies.

To enhance the predictive power of spatiotemporal data, advanced neural networks like Conv-LSTM models can be employed. Conv-LSTM neural networks combine convolutional layers for feature extraction with LSTM layers to capture temporal dynamics, offering a robust solution for stock market forecasting. The convolutional layers extract spatial features from the data, while the LSTM layers model the temporal dependencies, making the system well-suited for analysing spatiotemporal data in financial markets. This approach allows for a deeper understanding of regional influences and temporal trends that interact to affect market behaviour.

Moreover, the integration of Large Language Models (LLMs) into this framework further strengthens the analysis by processing unstructured textual data from financial news, social media, and economic reports. LLMs are designed to extract meaningful insights from unstructured text, adding a critical layer of context and sentiment analysis to the numerical data provided by spatiotemporal analysis. For instance, LLMs can analyse sentiment from financial news articles to gauge market sentiment and its potential impact on stock prices. This combination of spatial, temporal, and textual data provides a more holistic view of market dynamics, enabling investors to make more informed decisions.

In conclusion, the integration of spatiotemporal data with advanced neural networks, specifically Conv-LSTM models, offers a robust approach to stock market forecasting. Spatial data provides insights into the geographical distribution of market activities, while temporal data captures trends and patterns over time. The synergy of these data types allows for a comprehensive analysis of how regional and temporal factors interact to influence market behaviour. The incorporation of LLMs further enriches this analysis by adding context and sentiment from unstructured textual data, making the overall predictive model more sophisticated and accurate. This innovative approach demonstrates the power of combining multiple AI methodologies to tackle the complexity of financial forecasting, ultimately helping investors manage risks and make more strategic investment decisions.

## II. How Does a Conv-LSTM Work?

Long Short-Term Memory (LSTM) is a sophisticated type of recurrent neural network (RNN) architecture specifically designed to overcome the shortcomings of traditional RNNs, such as the vanishing and exploding gradient problems, which impede the learning of long-term dependencies. The LSTM architecture introduces a unique structure that includes a memory cell capable of maintaining information over long periods. This memory cell is regulated by three critical gates: the forget gate, input gate, and output gate. The forget gate determines which parts of the previous cell state should be retained or discarded, using a sigmoid activation function to scale the values between 0 and 1, thus deciding the extent of information preservation. The input gate controls the incorporation of new information into the cell state. It consists of two components: one that uses a sigmoid function to decide which values to update and another that uses a hyperbolic tangent (Tanh) function to generate potential new values to be added to the state. The cell state is then updated by combining the old cell state, modulated by the forget gate, with the new candidate values, modulated by the input gate. Finally, the output gate determines the next hidden state, which is used for the current output and transferred to the next time step, by applying a sigmoid function to decide which parts of the cell state to output and scaling this by the tanh of the updated cell state. This intricate mechanism allows LSTMs to effectively capture and utilize long-term dependencies, making them highly valuable for a range of applications involving sequential data. These include natural language processing tasks such as language modeling, text generation, machine translation, and speech recognition, as well as time series prediction tasks like stock market forecasting, weather prediction, and anomaly detection, and even control systems in robotics and automated processes. The ability of LSTMs to maintain context and handle long-term dependencies makes them a powerful tool for any task where understanding and processing sequences of data are essential.

Some notations:

$h_{t-1} = previous\ hidden\ state$

$x_t = current\ input$

$W_x = set\ of\ weights\ for\ respective\ gate(x)$

$b_x = bias\ term\ for\ respective\ gates(x)$

$C_t = cell\ state\ (updated\ by\ the\ forget\ and\ input\ gates.)$

The cell state:

$$\widetilde{C_t} = \tan h\ (\ W_c\ [\ h_{t-1}\ ,x_t\ ] + \ b_c$$

The Input gate $(i_t)$ of the LSTM architecture is responsible for determining what new information should be added to the cell during an iteration. This gate therefore helps regulate the flow of incoming information into the LSTM cell, ensuring that the model can selectively update its memory based on the current input and the previous hidden state.

$$i_t = \ \sigma(W_i \cdot [h_{t-1}, x_t] + \ b_i)$$

The Forget gate $(f_t)$ in LSTM architecture determines whether to keep the current value of memory or flush it. And to decide whether to which data to keep the forget date uses sigmoid function.

$$f_t = \ \sigma(W_f \ \cdot [\ h_{t-1}, x_t] + \ b_f$$

In an LSTM network, the forget gate and input gate work together to update the cell state, which acts as the memory of the network. The forget gate determines which information from the previous cell state $(C_{t-1})$ should be retained or discarded by multiplying it with a forget vector ($f_t$). If the outcome is 0, that information is dropped. The input gate then updates the cell state by adding new information from the input vector ($i_t$). This combination of retaining and updating gives the network a new cell state ($C_t$), which helps in making accurate predictions based on long-term dependencies.

$$C_t = \ f_t\ * C_{t-1} + i_t\ * \widetilde{C_t}$$

The Output gate $(o_t)$ control which pieces of information in the current cell state to output by assigning a value from

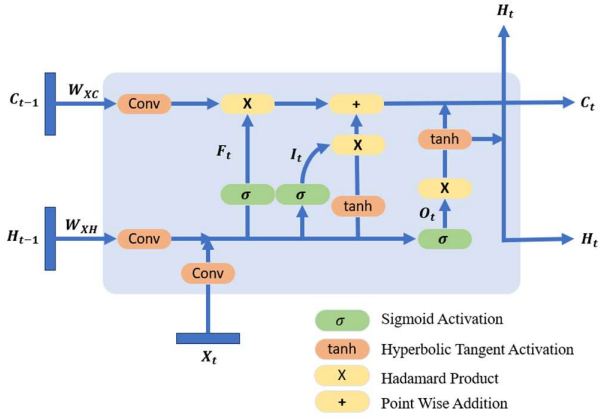

FIGURE 1: The internal implementation of a Conv-LSTM cell in a Conv-LSTM network. The cells take in the input from the previous state and after calculations (and applying convolution to all inputs) produce the output to be sent to the next state.

0 to 1 to the information, considering the previous and current states.

$$o_t = \sigma(W_o[h_{t-1}, x_t] + b_o)$$

In an LSTM cell, the final output ($h_t$) is determined by the Output Gate, which filters the cell state ($C_t$) to decide what information should be passed forward. The output gate applies a sigmoid function to the current input and previous hidden state, producing ($o_t$). This value is then multiplied with the tanh-activated cell state ($C_t$), resulting in the final output ($h_t$). This output is either used in the next LSTM cell or as the network's final prediction.

$$h_t = o_t * \tanh(C_t)$$

In conclusion, the final output of an LSTM cell, represented as ($h_t$), is a carefully filtered and modulated representation of the cell's memory state. By leveraging the Output Gate, the LSTM effectively determines which information is most relevant to pass forward, allowing the network to capture long-term dependencies and make more accurate predictions. This sophisticated process enables LSTMs to excel in tasks involving sequential data, such as time series forecasting, natural language processing, and stock market predictions.

## III. LARGE LANGUAGE MODELS

A Large Language Model (LLM) represents an advanced artificial intelligence system that excels in tasks involving natural language processing. These models are designed to comprehend and generate text that closely mimics human language by leveraging patterns and structures learned from extensive training datasets. Central to the architecture of LLMs is the transformer, a deep learning framework characterized by multiple layers of self-attention mechanisms. This architecture enables the model to evaluate the significance of various words or tokens in a sequence and to capture the intricate relationships between them. LLMs have been applied across a broad spectrum of domains. By incorporating additional supervised training data, these models can be fine-tuned for specific tasks, enabling them to excel in areas such as sentiment analysis, named entity recognition, or even complex problem-solving activities like playing chess.

Large Language Models (LLMs) follow a structured workflow that involves multiple stages, each critical to their performance in natural language processing tasks.

A. *Data Collection:* The initial step involves collecting large, diverse datasets from sources such as books, websites, and articles. This data forms the foundation for training the model, enabling it to develop a broad understanding of language patterns.
B. *Tokenization:* The textual data is then tokenized into smaller units (tokens), such as words or sub words, depending on the model. Tokenization allows the LLM to process text more efficiently and capture finer linguistic details.
C. *Pre-training:* In this phase, the model is trained to predict the next token in a sequence, using the transformer architecture. Pre-training is an unsupervised process in which the model learns grammar, semantics, and syntax by analysing vast amounts of data.
D. *Transformer Architecture:* LLMs are built on transformers, which use self-attention mechanisms to compute relationships between tokens. This allows the model to understand context and assign different weights to tokens based on their relevance in the sequence.
E. *Fine-tuning:* After pre-training, the model undergoes fine-tuning on task-specific datasets. This process adapts the LLM for specialized tasks, such as text classification, sentiment analysis, or question answering, using supervised learning.
F. *Inference:* Once trained, the model performs inference, generating predictions or text based on the input. During this stage, LLMs utilize their learned knowledge to produce contextually relevant outputs.
G. *Contextual Understanding and Beam Search:* LLMs excel at capturing long-range dependencies through self-attention mechanisms. For sequence generation, beam search is employed to generate the most likely sequence of tokens, ensuring coherent and contextually appropriate responses.
H. *Response Generation*: Finally, the model generates text by predicting subsequent tokens based on the input and previously generated tokens, producing fluent and human-like responses.

This workflow outlines the key stages in LLM training and usage, providing the foundation for their application in complex natural language tasks.

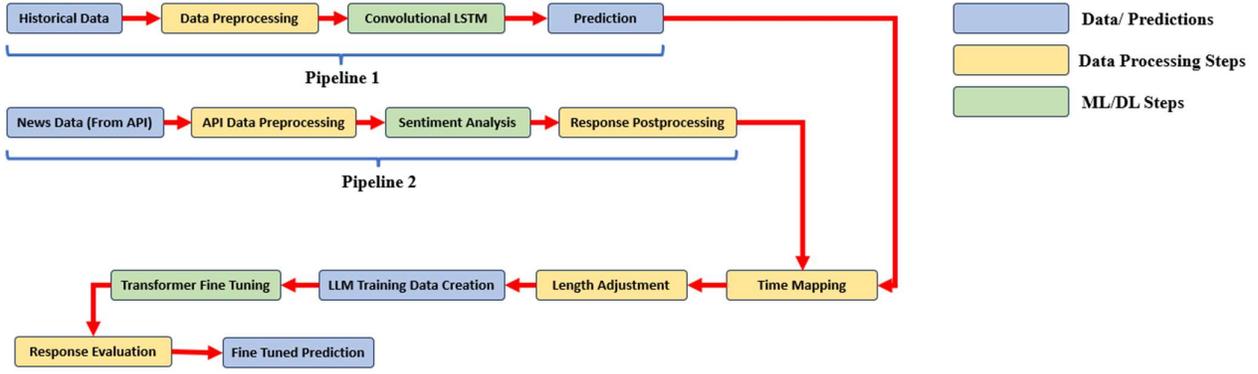

Figure 2: Machine Learning Workflow for fine-tuned forecasting using Convolutional LSTM and Transformer Model. Pipeline 1 processes historical data through data preprocessing, followed by Convolutional LSTM for prediction. Pipeline 2 integrates real-time news data via API, applying sentiment analysis and response postprocessing. Then the combined data from the pipelines form the training data for the transformer to produce the fine-tuned prediction.

## IV. PROBLEM WITH CONVENTIONAL FORECASTING

The conventional approach to stock forecasting leverages the powerful capability of the LSTM networks to capture the patterns in the temporal data provided to it and forecasting based on the same. However, what this model fails to do is capture the essential spatial data required to fine tune those predictions. As discussed in the previous section, the spatial data plays a very important role in determining the fate of the stock, e.g., a negative article from a reputed news channel may cause the value of the stock to plummet while a positive one may lead to unexpected rises in its valuation based on the intensity of the article. This, therefore, is very crucial for the investors at it might suggest them the right time for a profitable exit. Thus, a stock forecasting model purely based on how well a model can fit on the temporal data of the stock over a period of 5-10 years might end up not doing too well in real-world use cases. This is not ideal as it would be essential for our model to work well in both training and real-world scenarios. Therefore, it is much needed to incorporate spatial factors to fine tune the forecasting and achieve a state where the model can accurately predict the rise and fall in stock prices not only based on historical data but also on what is happening related to the stock in the world during the present times.

## V. THE PROPOSED SOLUTION

Analysis of spatial data components e.g. news articles etc. are required for fine tuning the LSTM prediction. The best models for such analysis tend to be the Large Language Models (or LLMs) which use the transformer architecture. Therefore, the proposed solution is a hierarchical model combining the temporal analysis capabilities of the initial conv-LSTM network with the spatial analysis provided by the LLM.

The hierarchical model consists of two layers, the first is the conv-LSTM layer that makes the initial forecast of the stock based on just the historical performance of the same. A separate pipeline can be built that collects all the news articles related to the stock over a given period. The news data can be fed into a pre-trained LLM (e.g. BERT) for sentiment analysis. The sentiment analysis is essential as it will assign a sentiment score between -1 (extremely negative) and 1 (extremely positive) to each of the articles based on the intensity of the articles. Then a weighted cumulative average score can be taken based on the influence/reputation of the news sources.

The second layer combines both the pipelines where the data from the conv-LSTM network is merged with the sentiment scores using time-mapping to create the training data for the next LLM. A pretrained LLM is taken and finetuned using the generated data. The input data consists of two parts – Input text and Target text. The Input text consists of the conv-LSTM predictions, coupled with the cumulative average sentiments using time-mapping. The target text will contain the actual value of the stock (fetched from the market) at that time. This will progressively fine tune the LLM model to help predict a value that will be closer to the actual market prices. Since sudden rises cannot be detected using time series analysis (i.e. LSTMs), the model depends on the news articles/other spatial features i.e., geographic area etc. to help predict a value closer to the target value at the outliers (unexpected highs and lows). Therefore, combining the LLM's ability to understand spatial features with the prediction made by Conv-LSTM using the historical data will help the model provide an overall holistic forecast of the stock and will perform well in real world scenarios.

## VI. WORKING OF THE MODEL

The first layer of the hierarchical LSTM-LLM architecture consists of feeding the historical data into the conv-LSTM to generate the naive forecast based on the historical patterns. The total data is split into sequences of optimal length based on testing.

*A. Data Preprocessing:*

The preprocessing of the data before loading it into the conv-LSTM is essential for the LSTM to function. The stock data is pre-processed to normalize and clean the dataset. This step ensures that the data is in a format suitable for feeding into the Conv-LSTM model. Preprocessing typically includes handling missing values, rescaling, and structuring the data to fit into a tensor format.

***1. Normalization:*** In this use case, the Z-score normalization is used. Unlike min-max scaling, which compresses values into a fixed range, Z-score normalization preserves the distribution of the data, making it suitable for algorithms that assume a Gaussian distribution (e.g., logistic regression, linear regression). Moreover, it improves the convergence of gradient-based optimizations, as features with vastly different scales can lead to erratic gradients.

$$x' = \frac{x - \mu}{\sigma}$$

Where x is the original value, $\mu$ is the mean of the dataset and $\sigma$ refers to the standard deviation of the same.

***2. Choosing the optimal group length:*** The next part of the preprocessing involves the splitting of the entire data into groups of optimal length for the LSTM to take in at once. This optimal length $'L'$ can be achieved by the following procedure:

$L_t$: Sequence length at iteration t

$P(L_t)$: Performance at sequence length $L_t$

$\Delta L_t$: Step size at iteration t, initially set to a large value.

$\eta$: Performance improvement threshold

$\alpha$: Reduction factor for $\Delta L$

$$L_{t+1} = L_t + \Delta L_t \text{ if } P(L_{t+1}) > P(L_t) + \eta$$

$$L_{t+1} = L_t - \Delta L_t \text{ if } P(L_{t+1}) < P(L_t) - \eta$$

$$\Delta L_{t+1} = \alpha \Delta L_t \text{ if } |P(L_{t+1}) - P(L_t)| < \eta$$

Stopping criteria: $|P(L_{t+1}) - P(L_t)| < \eta$ and $\Delta L_t < \varepsilon$. Here, $\epsilon$ is a very small value (e.g., 1) below which further refinements in sequence length are not useful. Therefore, $L_t$ obtained is now the optimal length of the group that can be used by the conv-LSTM for the next steps.

***3. Creating the training set:*** Now that the optimal length is found, the data can be grouped into groups of optimal length to capture the temporal dependencies of the data. The optimal length balances model complexity and computational efficiency. Here the sliding window approach is used, dividing time-series data into fixed-length windows, where each window contains a subset of past observations used as input to predict the next value(s) in the sequence. The windows can overlap, allowing the model to learn temporal dependencies and patterns from sequential data.

*B. Convolutional LSTM:*

In Conv-LSTM, the convolutional layers are responsible for capturing spatial features, which, in this case, can refer to patterns in multiple stock attributes or other external financial indicators. These spatial patterns are essential for understanding localized correlations between variables (e.g., stock prices, trading volume, and volatility) over time. Simultaneously, the LSTM layers capture the temporal dependencies in the time series stock data. LSTMs are known for their ability to retain important information over extended periods of time, making them highly effective at recognizing long-term patterns and trends in stock prices. This enables the Conv-LSTM to not only focus on short-term fluctuations but also account for long-term market behaviours, like seasonal trends or economic cycles.

Loss function of the same is:

$$\frac{1}{n} \sum_{i=1}^{n} (y_i - y_i')^2$$

i.e., the mean squared error (MSE) loss. However, a better alternative of the same is the 'Huber Loss' function that combines the strengths of MSE and MAE (mean absolute error) making it more robust to outliers.

$$Huber\ Loss = \begin{cases} \frac{1}{2}(y_i - y_i')^2; & if\ |y_i - y_i'| \leq \delta \\ \delta\left(|y_i - y_i'| - \frac{1}{2}\delta\right); & otherwise \end{cases}$$

Where, $\delta$ is a threshold that defines the point at which MSE transitions to MAE.

The conv-LSTM returns the predicted time series which can be then be carried forward to the next step.

*C. Processing and Tokenization of the News Data:*

***1. Data Fetch:*** The news data i.e., the news articles regarding a particular stock are fetched using an API (e.g. News API). The API call will return a JSON that can be parsed to get the relevant details.

***2. Data Processing:*** The names of the news/article website, the title of the articles and the body of the articles are taken together. The titles and bodies of the articles are then concatenated together and made ready for tokenization.

***3. Data Tokenization:*** The data needs to be tokenized to be accepted by the transformer in the next step. Therefore, tokenization takes place, removing any irrelevant characters and organizing the text data. The data is then fed into a natural language processing model (NLP) i.e., BERT in this case.

### D. Sentiment Analysis using BERT

Once tokenized, the cleaned and organized data is passed into a natural language processing (NLP) model, specifically BERT (Bidirectional Encoder Representations from Transformers) in this case. BERT uses its deep learning capabilities to understand the context of the words in the articles, capturing both the general sentiment and nuanced meanings, which is essential for further analysis or prediction tasks based on the stock news. Therefore, BERT is applied to the processed news data to analyse the sentiment of the text (positive, negative, or neutral). This sentiment score serves as an additional feature that can influence stock predictions, as positive news may indicate a rise in stock prices, while negative news can signal a fall.

### E. Response Postprocessing using Weighted Cumulative Score

The response generated by the BERT will contain a sentiment score mapped to each of the articles. Now, each day several articles are published regarding a particular stock. Therefore, the cumulative sentiment score needs to be calculated for the day or that time. Therefore, a weight to the respective article/news website name based on its influence/reputation is assigned.

***1. Setting the Sentiment Scores:*** The response from BERT for an article contains the tag "POSITIVE", "NEGATIVE" or "NEURAL". Therefore, the sentiment score is multiplied with (-1) if the tag is "NEGATIVE" and kept as is if it's one of the other two.

***2. Calculating the Weighted Cumulative Score:*** To calculate the weighted cumulative score each of the sentiment scores is multiplied with the respective weights of the articles/news. Then based on over what time the average sentiment score is to be calculated i.e., over the entire day, or the last hour, the average of the weighted sentiment scores is calculated and that is the weighted cumulative score for the stock over that period.

$$W_{cs} = \sum_{i=1}^{n} \frac{(w_i . x_i)}{w_i}$$

$w_i = weight\ of\ the\ i^{th}\ news\ article$

$x_i = Sentiment\ Score\ of\ the\ i^{th}\ newspaper$

$W_{cs} = Cumulative\ Sentiment\ Score\ of\ the\ stock$

Therefore, this weighted cumulative score represents the overall sentiment of the stock throughout the time interval and will be very helpful during the training of LLM.

### F. Time Mapping and Length Adjustment

The predicted series provided by the conv-LSTM gives the naïve forecast based on the historical data. The news data contains the weighted cumulative sentiment scores for the respective time intervals. Now, the data from both are mapped together using the time intervals. Each pair formed for each time interval contains the prediction from the conv-LSTM and the cumulative weighted sentiment score from the NLP model. Next, the length-adjustment takes place. This is an optional step. This is required when the shape of the LSTM prediction (i.e., the length) is not equal to that of the news data, i.e., in certain cases, enough news data might not be fetched to map the time frames for the entire historical data. Therefore, the length of the historical data is shortened to match that of the news data.

Therefore, the sentiment data is aligned with the corresponding time periods of the stock data, and shape adjustments are applied to ensure both datasets are synchronized, allowing for a more nuanced understanding of how news sentiment affects stock prices over time.

### G. Transformer Fine-Tuning

The combined dataset of time-series predictions and sentiment scores is used to train a Transformer model, specifically fine-tuning a T5 model. The T5 architecture is a sequence-to-sequence model well-suited for tasks that involve language generation and transformation, but here it is being fine-tuned for time series prediction. The data for fine-tuning the transformer is made from the combined spatiotemporal data obtained from combining both the data sources. The training data for the transformer will be of the form:

{

    "text": "LSTM prediction <conv-LSTM prediction> and sentiment score <weighted cumulative sentiment score>,

    "target": "actual target <target value>"

}

The transformer is trained over this training data and the response is taken to the final step.

### H. Response Evaluation

The predictions generated by the Transformer are evaluated for accuracy and reliability, ensuring that the model can effectively capture complex market dynamics involving both historical data and real-time sentiment analysis. After evaluation, the fine-tuned model outputs a final time series prediction. This prediction is informed by both the stock's historical behaviour (captured by the LSTM) and the real-time sentiment (captured by BERT).

### I. Overview of the pipeline:

In this system, the Conv-LSTM focuses on learning long-term dependencies and patterns in time-series stock data, while BERT analyses the sentiment of news articles to

TABLE I

ERROR METRICS COMPARISON FOR THE DEVELOPED HYBRID MODEL CALCULATED FOR CLOSE PRICE (NASDAQ: AAPL)

| Error Metrics | Machine Learning Model | |
|---|---|---|
| | Convolutional LSTM | Hybrid Model (conv-LSTM + LLM) |
| Mean Absolute Error (*MAE*) | 3.258327 | 1.605440 |
| Mean Squared Error (*MSE*) | 16.432614 | 4.190346 |
| Root Mean Squared Error (*RMSE*) | 4.053716 | 2.047034 |
| Mean Absolute Percentage Error (*MAPE*) | 1.448304 | 0.714751 |

TABLE II

ERROR METRICS COMPARISON FOR THE DEVELOPED HYBRID MODEL CALCULATED FOR CLOSE PRICE (NASDAQ: GOOG)

| Error Metrics | Machine Learning Model | |
|---|---|---|
| | Convolutional LSTM | Hybrid Model (conv-LSTM + LLM) |
| Mean Absolute Error (*MAE*) | 4.789342 | 1.955891 |
| Mean Squared Error (*MSE*) | 20.519253 | 6.248912 |
| Root Mean Squared Error (*RMSE*) | 4.529818 | 2.499782 |
| Mean Absolute Percentage Error (*MAPE*) | 2.902631 | 1.185388 |

gauge real-time market reactions. The two are integrated through time-based mapping and shape adjustment, and then fine-tuned using a Transformer model for enhanced prediction accuracy. The final output is a time series prediction that incorporates both historical trends and sentiment-driven fluctuations. This fusion of different AI models—Conv-LSTM for sequential data, BERT for textual sentiment analysis, and a Transformer for fine-tuning—creates a sophisticated, multi-dimensional approach to stock market forecasting.

## VII. RESULTS AND CONCLUSION

The dataset utilized in this research is a custom dataset that consists of historical stock data over the past four years, combined with related news articles from the same time. The stock data includes daily metrics such as closing prices, trading volumes, opening prices, and adjusted closing prices, capturing the stock's performance across various market conditions. In parallel, news articles were gathered using the NEWS API, which aggregates content from over c150,000 sources, including major media outlets and niche financial publications. These articles focus on events and developments relevant to the stock, such as financial earnings, product launches, and broader economic trends. This comprehensive dataset allowed us to analyse both quantitative financial data and qualitative news sentiment to assess their combined impact on stock behaviour.

The performance of the machine learning models was evaluated using several key error metrics, including Mean Absolute Error (MAE), Mean Squared Error (MSE), Root Mean Squared Error (RMSE), and Mean Absolute Percentage Error (MAPE). The results demonstrate that the Hybrid Model (combining Convolutional LSTM and LLM) significantly outperformed the standalone Convolutional LSTM model across all metrics. The improvement in performance suggests a direct relationship of the stock's performance with the news data that was provided to it. This implies that while using only historical data can already result in accurate predictions since the temporal trends are captured by the LSTM models, we can further enhance the accuracy of the model by incorporating the spatial data analysis related to the stock since it will help establish a relationship between the spatial features obtained during training and thus help in improving the overall accuracy of the model.

In addition to stock price prediction, the hybrid approach of combining quantitative time-series data with qualitative contextual data, such as news sentiment, has broad potential applications in other fields. For instance, in the healthcare industry, predictive models could integrate historical patient data with medical literature or news articles on emerging treatments to forecast patient outcomes or disease trends more accurately. Similarly, in supply chain management, models could use historical inventory data alongside news reports on global logistics, economic policies, or environmental conditions to predict potential disruptions or optimize stock levels. The fusion of temporal and contextual information, as demonstrated in this research, opens new possibilities for making more informed and accurate predictions across a wide range of domains, where external factors play a critical role in determining outcomes. This approach not only enhances prediction accuracy but also provides more comprehensive insights for decision-makers in various industries.